 \pdfoutput=1
\documentclass[%
  prb,%
  letterpaper,%
  twocolumn,%
  showpacs,%
  floatfix,%
  superscriptaddress%
]{revtex4}

\RequirePackage{graphicx,amsmath,amssymb,bm}
\RequirePackage{hyperref}
\hypersetup{%
  breaklinks = {true},
  citecolor = {blue},
  colorlinks = {true},
  linkcolor = {red},
  pdfauthor = {\textcopyright\ D. A. Bahamon},
  pdfcreator = {\LaTeX\ and \flqq hyperref\frqq},
  pdffitwindow = {true},
  pdfmenubar = {true},
  pdfpagelayout = {SinglePage},
  pdfstartview = {Fit},
  pdftoolbar = {true},
  plainpages = {false},
}

\newcommand{\Fref}[1]{Fig.~\ref{#1}}
\newcommand{\Eqref}[1]{Eq.~(\ref{#1})}
\renewcommand{\eqref}[1]{eq.~(\ref{#1})}

\RequirePackage[usenames,dvipsnames]{xcolor}
\usepackage{soul}

\begin{document}

\title{Graphene kirigami as a platform for stretchable and tunable quantum dot arrays}

\author{D. A. Bahamon%
}
\affiliation{MackGraphe\,--\,Graphene and Nano-Materials Research Center, 
Mackenzie Presbyterian University, Rua da Consola\c{c}\~{a}o 896, 01302-907, 
S\~{a}o Paulo, SP, Brazil} 

\author{Zenan~Qi%
}
\affiliation{Department of Mechanical Engineering, Boston University, 
Boston, MA 02215}

\author{Harold~S.~Park%
}
\affiliation{Department of Mechanical Engineering, Boston University, 
Boston, MA 02215}

\author{Vitor~M.~Pereira%
}
\affiliation{Centre for Advanced 2D Materials \& Graphene Research Centre
National University of Singapore, 6 Science Drive 2, Singapore 117546}
\affiliation{Department of Physics, National University of Singapore, 
2 Science Drive 3, Singapore 117542}

\author{David~K.~Campbell%
}
\affiliation{Department of Physics, Boston University, 
590 Commonwealth Ave, Boston, MA 02215, USA}

\date{\today}

\begin{abstract}
The quantum transport properties of a graphene kirigami similar to those studied in recent experiments are calculated in the regime of elastic, reversible deformations. Our results show that, at low electronic densities, the conductance profile of such structures replicates that of a system of coupled quantum dots, characterized by a sequence of minibands and stop-gaps. The conductance and I-V curves have different characteristics in the distinct stages of deformation that characterize the elongation of these structures. Notably, the effective coupling between localized states is strongly reduced in the small elongation stage but revived at large elongations that allow the reestablishment of resonant tunneling across the kirigami. This provides an interesting example of interplay between geometry, strain, spatial confinement and electronic transport. The alternating miniband and stop-gap structure in the transmission leads to I-V characteristics with negative differential conductance in well defined energy/doping ranges. These effects should be stable in a realistic scenario that includes edge roughness and Coulomb interactions, as these are expected to further promote localization of states at low energies in narrow segments of graphene nanostructures.
\end{abstract}



\maketitle

The development of advanced microfabrication techniques in the final two decades of the last century allowed the creation of two-dimensional electron gases (2DEG) at the gate oxide-semiconductor interface of heterostructures \cite{Beenakker19911}. Electrons in these structures can be considered as free in the directions parallel to the interface but strongly confined in the transverse direction. Advances in the design of metallic gates with specific patterns on the heterostructure surface enabled the on-demand depletion of electrons in predetermined spatial regions, thus allowing the experimental study of quantum point contacts\cite{PhysRevLett.60.848,PhysRevB.38.3625}, quantum wires \cite{PhysRevLett.59.3011}, constrictions\cite{Wuconstriction} and quantum dots\cite{KOUWENHOVEN1990290}. Electron transport experiments in patterned periodic metallic gate structures\cite{PhysRevLett.65.361} show the formation of minibands and transport gaps and, by fine-tuning the gate voltages, the coupling between the periodic quantum dots can be tuned, revealing the formation of confined states at individual sections\cite{PhysRevLett.65.361}. 

Today, truly two-dimensional electronic systems are routinely achieved in 
atomically thin materials, of which graphene is the best known and most widely 
studied example \cite{Rozhkov:2011jk}. Unlike traditional 2DEG built at semiconductor-oxide 
interfaces, the electronic system in these atomically-thin crystals is directly 
exposed and thus much more amenable to external control. In graphene, however, 
the relativistic-like electronic dispersion strongly reduces the effectiveness 
of electrostatic gating as a means to establish confining electrostatic 
potentials due to the Klein tunneling effect \cite{Katsnelson:2006ul}. 
Consequently, quantum dots and other constrained structures for the experimental 
study of quantum transport characteristics have been traditionally fabricated by 
direct, and permanent, etching on the material: lithographic techniques in 
graphene have been used to confine electrons in quantum dots 
\cite{Ponomarenko:2008dq,PhysRevB.77.115423,PhysRevB.85.165446,arXiv:1601.00986}, nanoribbons \cite{Tapaszto:2008db,PhysRevB.80.121407} and  junctions 
\cite{Tayari:2015rw}. 

\begin{figure}[tb]
\vspace{-0.2cm}
\centerline{\includegraphics[width=0.45\textwidth]{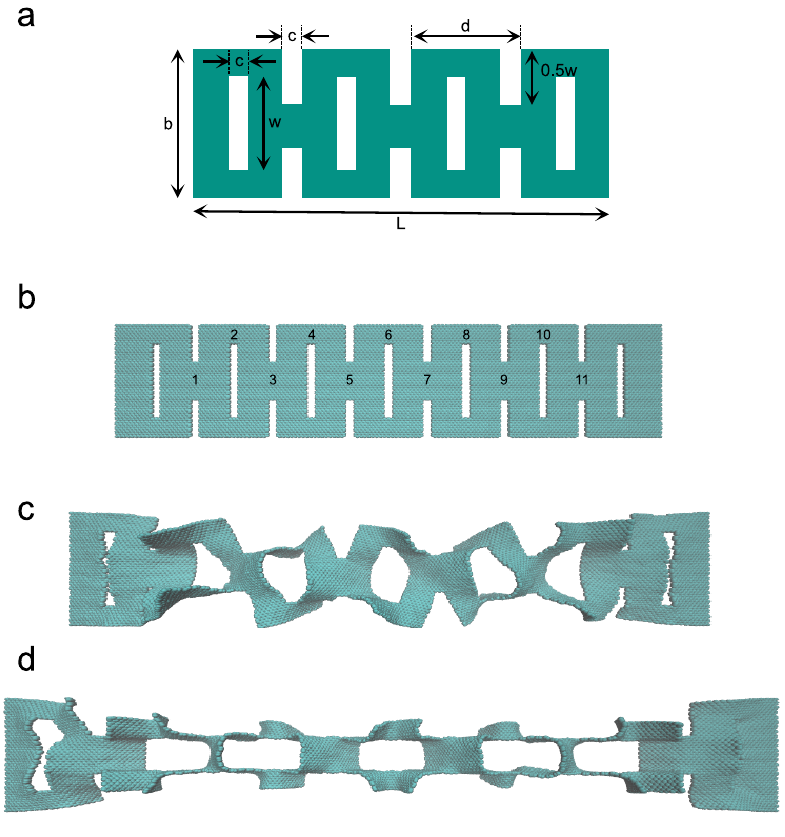}}
\vspace{-0.2cm}
\caption{ (a) Pattern schematics of the graphene kirigami indicating the most relevant geometric parameters. Snapshots of the actual kirigami used in the calculations for deformations of (b) 0\%, (c) 15.5\% and (d) 34.7\%. The figures were generated by VMD\cite{Humphrey:1996fk}.  }
 \label{fig:structure}
\end{figure}

An interesting and different direction has recently been pursued by employing lithography to extend the range of elastic deformation that can be sustained by a graphene by patterning nanomeshes out of large graphene sheets \cite{Zhu_nanomesh}. Such structures, inspired by the Japanese art of cutting paper called kirigami, have been designed and tested in recent experiments \cite{Blees:2015fk} that establish their mechanical robustness and extremely high elongation limits ($\sim 240$\%) compared to uncut graphene. In a previous study of the elastic and mechanical characteristics of graphene kirigami, some of us showed that their stretchability limit and effective Young's modulus can be characterized (and customized) in terms of two geometric parameters \cite{PhysRevB.90.245437}: $\alpha$, the ratio of the overlapping cut length to the kirigami length, and $\beta$, the ratio of the overlapping width to the kirigami length. With reference to \Fref{fig:structure}a, these parameters are $\alpha = (w-0.5b)/L$ and $\beta = (0.5d-c)/L$. 

The ability currently demonstrated to experimentally design graphene kirigami capable of ultra high elastic deformations raises the question of how the electronic states and the flow of current within the kirigami are, or can be, modified under deformation. In particular, since any kirigami always involves a number of bends, indentations, and narrow regions (henceforth, ``constrictions''), it can naturally harbor a number of localized states at low energies (below the threshold for electronic transmission) as a result of either the geometry alone \cite{PhysRevB.83.155450}, or geometry combined with disorder \cite{PhysRevB.80.155415} and/or Coulomb interactions \cite{Todd2008Quantum,PhysRevLett.104.056801}. The experimental observation of Coulomb blockade in graphene nanoribbons and constrictions \cite{Sols2007Coulomb,Todd2008Quantum,Stampfer2009Energy,Gallagher2010Disorderinduced} indicates that such states are expected to be prevalent at low energies in ``papercut'' graphene devices and that disorder can promote or further stabilize them.

A segment of suitably patterned graphene hosting such states defines a quantum dot, and its periodic repetition would define an array of coupled quantum dots, analogously to what has been achieved in semiconductor structures through comb-shaped arrays of side gates \cite{PhysRevLett.65.361,PhysRevB.43.12082,PhysRevB.45.6652}. The fact that graphene kirigami of possibly any desired shape can sustain repeated stretching cycles \cite{Blees:2015fk} suggests that one might be able to design kirigami suitable for the study of coupled quantum dots, with the advantage that the inter-dot coupling responds directly to the deformed state of the structure. The vast range of stretchability limits and effective Young moduli that these structures can be designed with, combined with the proven mechanical resilience of graphene at the nanoscale, hint at the possibility of designing versatile electro-mechanical devices based on this concept of stretchable graphene quantum dot arrays whose current response can be strongly sensitive to the inter-dot coupling.

That is precisely the problem we address in this work, concentrating only in the geometric aspects that contribute to localization of electronic states at certain portions of the kirigami. We confirm that such states define effective quantum dots, which result in a characteristic profile of the conductance at low energies defined by a sequence of resonant mini-bands and stop-gaps, see \Fref{fig:figTE}a. This is analogous to the conductance of periodic split-gate semiconductor heterostructures \cite{PhysRevLett.65.361,PhysRevB.41.12350,PhysRevB.43.12082}, where each miniband consists of a set of resonant tunneling peaks. At low energies, the conductance profile is seen to be very sensitive to the deformation, and we analyze this response in terms of the interplay between variations in inter-dot coupling and strain barriers that develop under stretching. 

Our study combines information of the local atomic displacements obtained from molecular dynamics (MD) simulations of deformed kirigami with quantum transport calculations to assess the conductance characteristics at different stages of deformation \cite{C5NR03393D}.  At the structural level, the initial stage (stage 1) of the end-to-end longitudinal deformation is characterized by a very small effective Young's modulus and an essentially negligible average stretching of the inter-atomic bonds because, in this early stage, elongation occurs mostly through bending and twisting of the structure in three-dimensional space \cite{PhysRevB.90.245437}. Despite the low in-plane stretching that occurs during this stage, the conductance and the current are significantly reduced, and the resonant features disappear. However, further elongation into a second stage of deformation where there is substantial bond stretching leads to the revival of the resonant features, and the restoration of the magnitudes of conductance to the same levels observed in the undistorted kirigami. These changes are consistently understood as a consequence of transport at low energies being dominated by resonant tunneling between states localized at specific portions of each periodic unit of the kirigami: elongation during stage 1 of deformation perturbs the coupling (overlap) between these states and considerably degrades the conductance as a result of hopping disorder that weakens the super-periodicity of the kirigami structure; further stretching into stage 2 results in strong and localized strain barriers that re-establish the super-periodicity, favor electron localization, and hence revive resonant transmission. 
The terminology ``strain barrier'' is employed here deliberately to emphasize the direct impact that deformation hot-spots have in the electronic behavior of the system at high elongations: as discussed below, the development of localized and periodic regions with extreme bond stretching reinforces the geometry-induced confinement of electronic states within each repeating unit of the kirigami.

\section{Methodology}

Our representative kirigami is obtained by cutting out a graphene nanoribbon of width $b$ according to the pattern shown in \Fref{fig:structure}a. The rectangular interior holes are defined by the height $w$ and width $c$, while the connecting necks have length $c$ and a cut starting from the outer edge of length $0.5w$. The longitudinal period is $d$ and $L$ defines the total length of the system. We shall concentrate our discussion in the kirigami shown schematically in \Fref{fig:structure}b: our actual system contains 11408 atoms with the geometrical parameters  $b \approx 10$\,nm, $w \approx 6.8$\,nm, $c \approx 0.7$\,nm, $d=4.8$\,nm and $L \approx 34$\,nm. For definiteness, we base our discussion on this specific structure where the graphene lattice is oriented so that all horizontal edges are zig-zag edges. We also do not include any disorder or edge roughness at this stage. The results, however, are general and should hold when these two restrictions are relaxed because \emph{the only key physical ingredient is the existence of localized states defining a local quantum dot} at specific constrictions or bends, and these can be stabilized by different geometries, with or without disorder \cite{PhysRevB.80.155415}. For illustration, we show that explicitly by analyzing a different lattice orientation in the discussion session.

This study is divided in three stages that aim to determine the electronic 
transport properties of the kirigami under longitudinal tension. The deformed 
structures are first obtained obtained with molecular dynamics (MD) simulations 
of the finite kirigami. We used the Sandia open-source code LAMMPS 
\cite{lammps,Plimpton:1995ys} with the AIREBO 
\cite{ZhaoMD,StuartMD,0957-4484-21-26-265702} potential to describe the C-C 
interactions: the cutoff radii are 2\,\AA~for the REBO term and 6.8\,\AA~for the 
Lennard-Jones term in the AIREBO potential. The explicit position of each carbon 
atom in the deformed structure is then used to build a $\pi$-band tight-binding 
Hamiltonian, $H =   \sum_{<i,j>}t_{ij} (c_{i}^{\dagger} c_{j} + 
c_{j}^{\dagger} c_{i})$, for the distorted kirigami. In this Hamiltonian $c_{i}$ denotes the annihilation operator on site $i$ and $t_{ij}$ represents the hopping amplitude between nearest neighbor sites $i$ and $j$. The stretching, the compression and the rotation of the C-C bonds created by the tensile load are taken into account in this effective Hamiltonian by means of a position-dependent hopping parameter that reflects the overlap between two arbitrarily oriented $p_z$ orbitals\cite{C5NR03393D}:
\begin{multline}
  t_{ij}(d)=V_{pp\pi}(d_{ij})\, \hat{n}_i \cdot \hat{n}_j \\
  + \Bigl[V_{pp\sigma}(d_{ij})-V_{pp\pi}(d_{ij})\Bigr] \frac{(\hat{n}_i \cdot 
  \vec{d}_{ij})
  (\hat{n}_j \cdot \vec{d}_{ij}) }{d^2_{ij}}
  .
  \label{eq:tij}
\end{multline}
Here, $\hat{n}_i$ is the unit normal vector to the surface at site $i$, $\vec{d}_{ij}$ is the vector connecting sites $i$ and $j$, and $V_{pp\sigma}$(d) and $V_{pp\pi}$(d) are the Slater-Koster integrals. The presence of the local normals accounts for the relative rotation of neighboring orbitals, whereas the effect of bond stretching is captured by the distance dependence of the Slater-Koster parameters, which we assumed to vary as \cite{PereiraPRB2009,QiPRB2014}:
\begin{align}
  \label{eq:SKp1}
  V_{pp\pi}(d_{ij}) & = t_0 \, e^{-\beta(d_{ij}/a-1)}, \\
  \label{eq:SKp2}
  V_{pp\sigma}(d_{ij}) & = 1.7 \, V_{pp\pi}(d_{ij}),
\end{align}
where $t_0=2.7$\,eV, $a\simeq 1.42$\,\AA\ is the equilibrium C-C bond length in graphene, and $\beta = 3.37$ captures the exponential decrease in the hopping with inter-atomic distance. All these quantities are easily calculated using the atomic positions provided by the MD simulations.

Subsequently, in order to inject charge into the system we couple two undeformed semi-infinite graphene nanoribbons to the left and right edges of the kirigami, the ``contacts'' being of the same width as the kirigami. To guarantee that the properties observed  are those of the kirigami and not of the deformed contact or the contact-kirigami interface, we keep the contacts undeformed during all stages of deformation.  In terms of Green's functions, the conductance in the Landauer-B\"uttiker formalism can be written as\cite{Caroli,Datta,Jauho}
\begin{equation}
  G=\frac{2e^2}{h} T(E) = \frac{2e^2}{h} \text{Tr}[\Gamma_RG^r\Gamma_LG^a]  
  ,
\end{equation}
where $G^r=[G^a]^{\dagger}=[E+i\eta-H-\Sigma_L-\Sigma_R]^{-1}$ is the retarded (advanced) Green's function, the coupling between the contacts and the central region is represented by $\Gamma_{L(R)}=i[\Sigma_{L(R)}-\Sigma_{L(R)}^{\dagger}]$, and $\Sigma_{L(R)}$ is the self-energy of left(right) contact. 
To correlate the conductance features with the real-space distribution of the electronic states, we map the local density of states (LDOS) at a given site $i$ directly from the local Green's function according to the identity $\rho_{ii}=-\text{Im}[G^r(\vec{r_i},\vec{r_i},E)] / \pi$. The $I-V$ curves of the device, where $I$ is the total current as a function of the applied bias voltage $V$, are calculated from the transmission function $T(E,V)$ as\cite{Datta} %
\begin{align}
  I(V) = \frac{2e}{h} \int^{+\infty}_{-\infty} T(E,V) \left[ f_L(E) - f_R(E)\right] dE
  ,
  \label{eq:Itot}
\end{align}
where $f_{L(R)}(E)$ is the Fermi distribution of the left (right) contact.

\section{Conductance under deformation}

\begin{figure}[tb]
\vspace{-0.2cm}
\centerline{\includegraphics[width=0.45\textwidth]{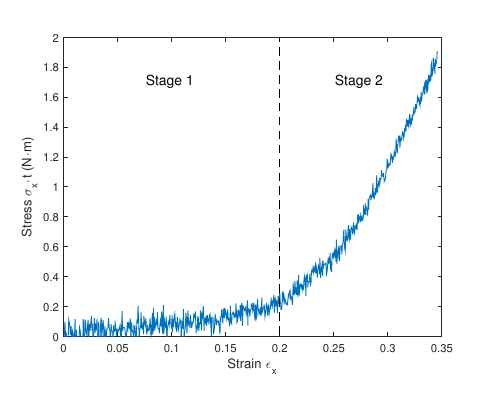}}
\vspace{-0.2cm}
\caption{ Stress-Strain curve for the kirigami, showing the first two stages corresponding to the elastic and reversible deformations.  }
 \label{fig:stressStrain}
\end{figure}

For a representative illustration showing the behavior in the two stages of deformation, we chose the kirigami structure shown in \Fref{fig:structure}b that has a small number of cuts per repeating unit and a very high stretchability, up to 65\,\%. In order to facilitate the presence of localized states within each segment of the kirigami even in the absence of any deformation, we chose to orient the underlying graphene lattice so that the longitudinal cuts are along a zig-zag direction, as the internal mini-zigzag edges so defined are expected to naturally support localized states \cite{zigzagedges} (as pointed out above, this is not a limitation).

A kirigami such as this one has four deformation stages \cite{PhysRevB.90.245437}: (i) elongation with bending and twisting, but very little in-plane stress, (ii) elongation with stress, (iii) yielding, and (iv) fracture. We restrict our analysis to the first two, where deformations are elastic and reversible, and whose stress-strain characteristic is shown in \Fref{fig:stressStrain}. During the first stage (in this particular structure that corresponds to total deformations below $\approx 20$\,\%) horizontal and vertical segments twist and rotate and, as a result of this excursion of the graphene sheet into the third dimension, the kirigami elongates without significant modification of the average C-C bond length, except for very localized strain hot-spots at the corners of the connecting elements \cite{PhysRevB.90.245437}. A representative sample of a kirigami in this stage is shown in \Fref{fig:structure}c for a total deformation of 15.5\,\%. With further increase in the tensile load, the kirigami is not capable of accommodating higher elongations only by twisting, and this triggers the onset of stage 2 (here in the range 20-40\,\%), where further deformation occurs through stretching of the carbon bonds. One important consequence of the existence of these two regimes is that the effective Young's modulus is much lower in stage 1, $E\simeq 0.69$ N/m, 
(where it is essentially determined by the very small bending stiffness of monolayer graphene) than in stage 2 $E\simeq15.08$ N/m.
This makes mechanical manipulation of kirigami structures experimentally possible and easy in stage 1 \cite{Blees:2015fk}.

\begin{figure}[tb]
\vspace{-0.2cm}
\centerline{\includegraphics[width=0.45\textwidth]{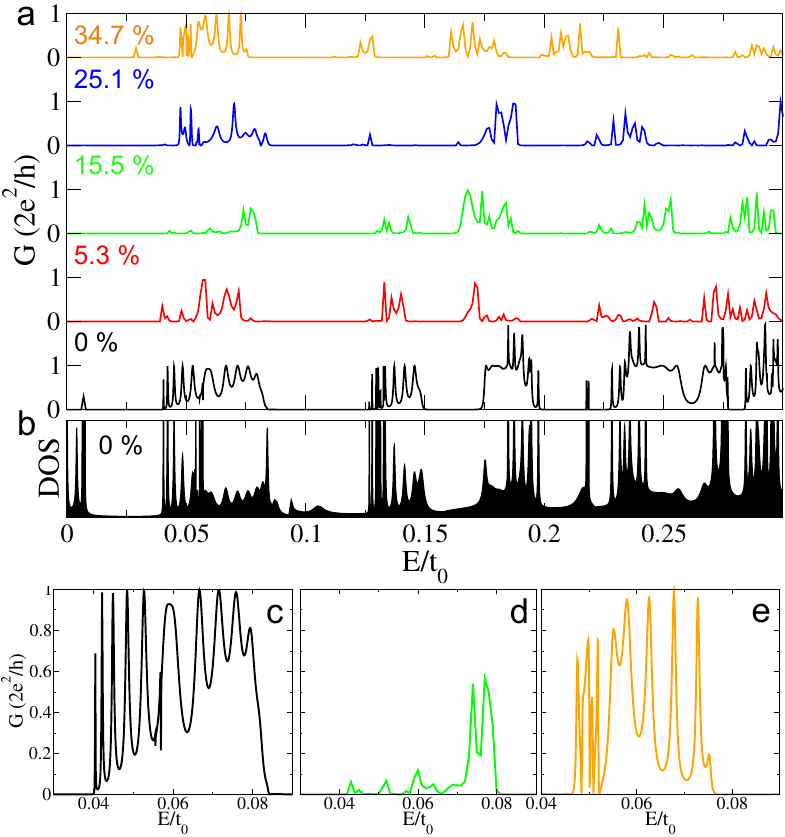}}
\vspace{-0.2cm}
\caption{ (a) Conductance of the kirigami for different values of deformation. (b) DOS of the undeformed structure. The lower row of panels contains a magnification of the energy range comprising the first group of resonances ($0.03 < E/t_0 < 0.09$) for the undeformed kirigami (c), kirigami deformed by 15.5\,\% (d), and deformed by 34.7\,\% (e).}
 \label{fig:figTE}
\end{figure}

We followed the evolution of conductance at low energies for different strain values within the reversible and elastic region and summarize the results in \Fref{fig:figTE}a for  deformations of 0\,\%, 5.3\,\%, 15.5\,\%, 25.1\,\%  and 34.7\,\%. The conductance profile of the reference (undeformed) structure displays resonant transmission within well defined bands (mini-bands) of width $\approx 0.05\,t_0$. We can see in the DOS plotted in \Fref{fig:figTE}b for the undeformed structure that each miniband arises from the clustering of a finite number of states. At those energies transmission occurs through resonant tunneling, and is entirely suppressed  otherwise, strikingly different from the conductance profile expected for a graphene nanoribbon \cite{PhysRevB.41.12350,PhysRevB.83.155450}. 
In an infinite kirigami the stop-gaps that separate the minibands arise from the folding of the 1D Brillouin zone of the ideal (pre-cut) graphene ribbon as a result of the new superlattice defined by the kirigami, with the expected opening of spectral gaps at the edge of the reduced zone. In our finite kirigami, \Fref{fig:figTE}b shows that the DOS is likewise strongly suppressed at the minband edges but does not reach zero in the stop-gaps. This contrasts with the sharp changes in the conductance [\Fref{fig:figTE}a]: finite (resonant) within a miniband, and clearly zero in the stop-gaps. This is a clear sign that states contributing to the DOS in the stop-gap regions are spatially localized, and unable to hybridize to form a tunneling pathway that spans the entire length of the system. Therefore, the robust stop-gap structure in the transmission is not a result of spectral characteristics of the kirigami alone.

\begin{figure}[tb]
\vspace{-0.2cm}
\centerline{\includegraphics[width=0.38\textwidth]{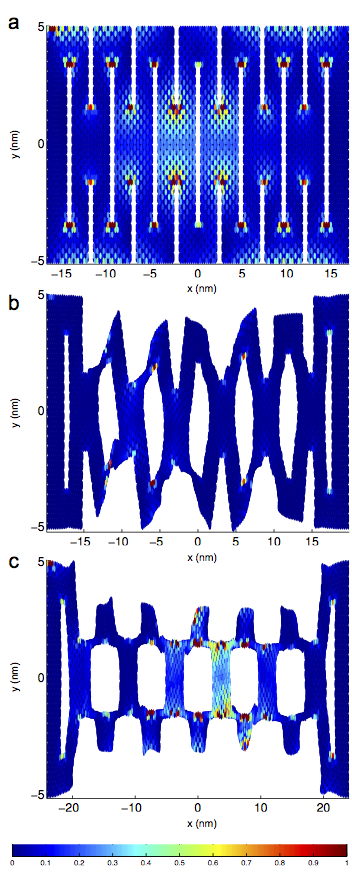}}
\vspace{-0.2cm}
\caption{Normalized LDOS for the kirigami (a) undeformed at $E = 0.067t_0$, (b) deformed by 15.5\,\% at $E = 0.074t_0$, and (c) deformed by 34.7\,\% at $E = 0.073t_0$.}
\label{fig:figLDOS}
\end{figure}

\begin{figure}[bt]
\centerline{\includegraphics[width=0.5\textwidth]{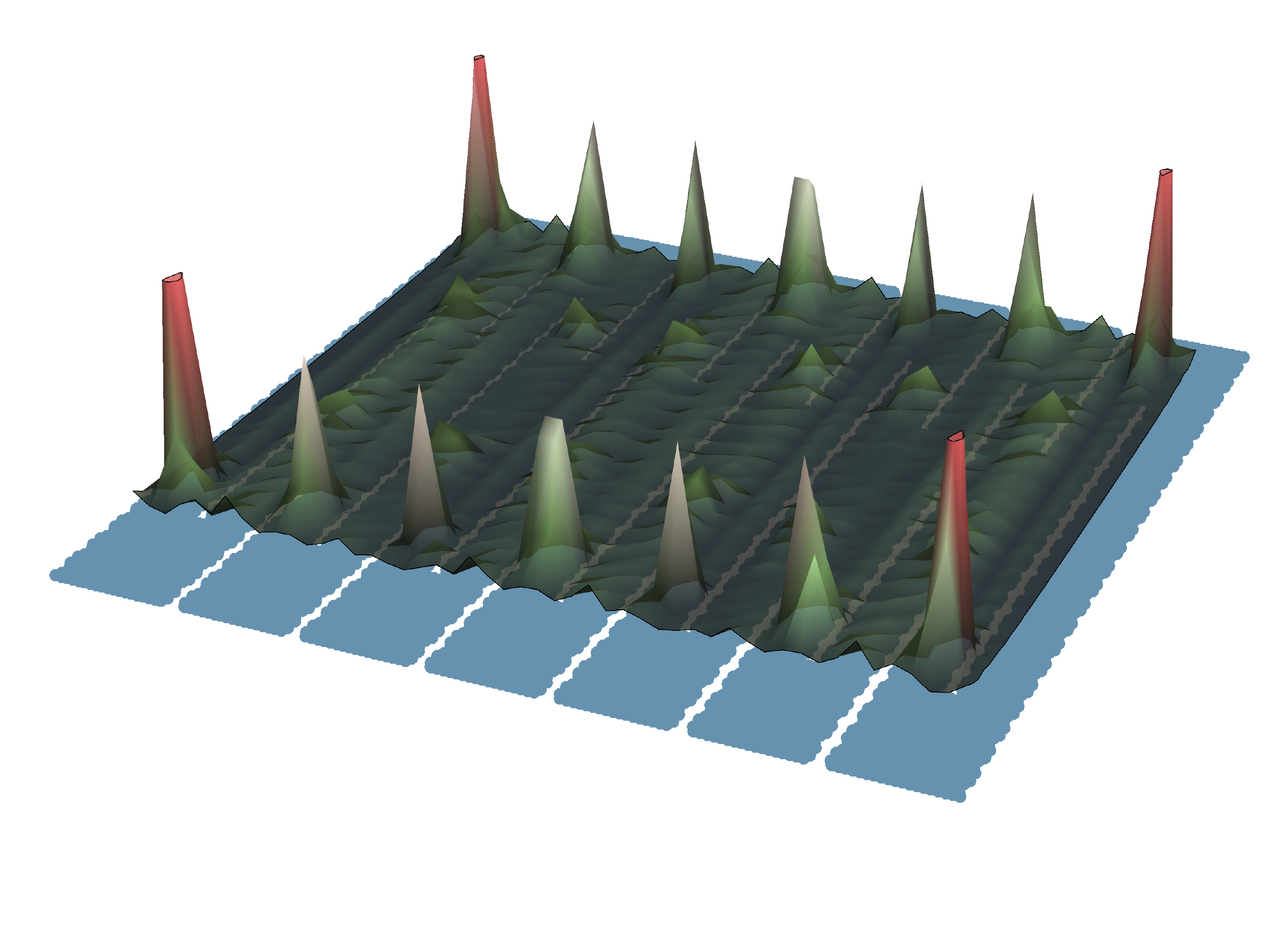}}
\caption{LDOS at one of the transmission peaks ($E=0.044t_0$) that contributes 
to the lowest transmission miniband of the undeformed kirigami, as discussed in 
the main text. The corresponding kirigami structure is shown underneath the 
density density plot. 
}
 \label{fig:ldos_3D}
\end{figure}

To be specific, let us analyze the close-up plot shown in \Fref{fig:figTE}c that 
captures the energy interval $0.03 < E/t_0 < 0.09$. A mapping of the LDOS at any 
of the resonances shown in this panel confirms that the electronic density 
associated with a resonance is strongly localized around the internal 
longitudinal strips. \Fref{fig:figLDOS}a shows a representative example of the 
``LDOS hot spots'' (in red) appearing at those regions, and a good 
overlap/coupling with the external contacts. Another snapshot of the LDOS at a different transmission resonance is shown in \Fref{fig:ldos_3D}.
The existence of the mini-bands of resonant transmission is therefore due to the 
presence of localized states within each unit of the kirigami, which are 
localized both transversely and longitudinally by the combined effect of the 
internal zig-zag edges and the finite extent of each segment of the 
kirigami\cite{PhysRevB.83.155450,ZhangBentQD,Wu:2010eu,Faria:2015pd, 
PhysRevLett.65.361,PhysRevB.41.12350,PhysRevB.43.12082}. At the lowest energies, 
transmission across the entire system is assisted by tunneling through these 
localized states that, hence, play a role similar to that of an array of coupled 
quantum dots that can enable resonant transmission across the entire system at 
(and only at) well defined energies. The width of each miniband is determined by 
the overlap, $\gamma$, between localized states, and the number of 
resonant peaks in a mini-band, $N$, counts the number of isolated quantum dots 
\cite{PhysRevB.41.12350,PhysRevB.45.6652}.
Our geometry contains the 11 internal channels labeled 1--11 in \Fref{fig:structure}b that define 11 effective quantum dots. Correspondingly, each transmission miniband has a fine structure consisting of 11 peaks as can be seen in \Fref{fig:figTE}c (the outermost two directly in contact with the metallic leads do not define quantum dots because backscattered electrons at these sections are completely absorbed by the contacts).

A quantitative estimate of the inter-dot overlap can be obtained assuming a 
one-dimensional tight-binding model, according to which the energy levels within 
one miniband should appear at positions
\begin{equation}
  E_n = -2\gamma\, \cos\Bigl(\frac{n\pi}{N+1}\Bigr)
  ,\qquad n=1,2,\dots,N,
\end{equation}
relative to the center of the miniband. The width of a miniband is given by 
$\Delta E = E_N-E_1$. The data in \Fref{fig:figTE}(c) allow us to estimate 
$\Delta E \approx 0.04 t_0$ for the undeformed kirigami, which corresponds to 
$\gamma \approx 0.01\,t_0$. Recalling that $t_0$ is the hopping between nearest 
neighboring carbon atoms in undeformed graphene [\Eqref{eq:SKp1}], this shows 
an inter-dot overlap two orders of magnitude smaller than the hopping in the 
underlying graphene lattice. Such a small value of $\gamma$ is natural given 
the large spatial separation ($\sim d/2 = 2.4$\,nm, cf. \Fref{fig:structure}a) 
between each pair of hybridized localized states that we see in 
\Fref{fig:ldos_3D}.

Under tensile load in stage 1, the kirigami elongates and becomes distorted, which significantly perturbs the overlap between the localized states. That is the conclusion that follows from the progressive disappearance of the resonant mini-bands under deformation that we can see in \Fref{fig:figTE}a for deformations of 5.3 and 15.5\,\%. It is notable that, even though in stage 1 there is very little change in the \emph{average} C-C bond length, the regions that are most affected by that, and which are more strongly bent, are those in the vicinity of the localized states of the undeformed kirigami. It is, therefore, not surprising that the conductance assisted by resonant tunneling through these states can be significantly modified in this stage since the perturbations to the hopping can be significant in precisely the regions more critical for the overlap between neighboring localized states. For an elongation of 15.5\,\%, \Fref{fig:figTE}a shows that the miniband structure disappears and the system behaves largely as an insulator through most of the energy range shown. The LDOS associated with the few remaining weak conductance peaks reflects the less effective coupling between the localized states (see \Fref{fig:figLDOS}a). In the analogy introduced above, in this stage of deformation we have the equivalent of a device consisting of distinct and asymmetrically coupled quantum dots, for which the energy of transmission resonances is predominantly determined by the energy levels and coupling of individual dots \cite{JoeAsymmQD,PhysRevB.49.14736}. 

\begin{figure}[tb]
\vspace{-0.2cm}
\centerline{\includegraphics[width=0.45\textwidth]{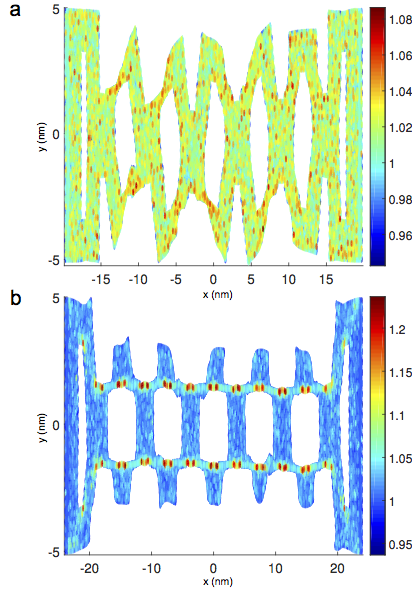}}
\vspace{-0.2cm}
\caption{Normalized C-C bond length for the kirigami deformed by 15.5\,\% (panel a, stage 1) and 34.7\,\% (panel b, stage 2), where the normalization is by the undeformed C-C bond length.}
\label{fig:figSIV} 
\end{figure}

\begin{figure}[tb]
\vspace{-0.2cm}
\centerline{\includegraphics[width=0.45\textwidth]{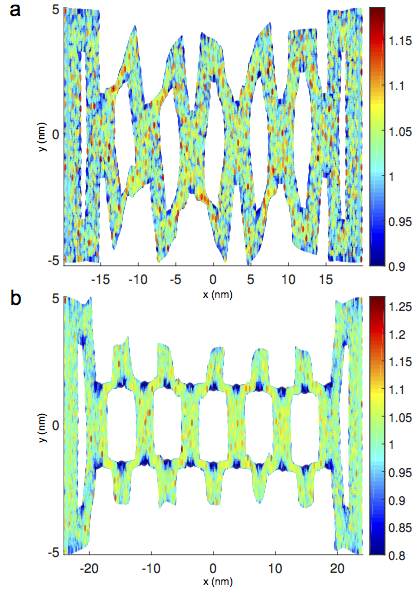}}
\vspace{-0.2cm}
\caption{The distribution of the nearest-neighbor hopping amplitudes $t_{ij} / t_0$ calculated according to \eqref{eq:tij} for the same structures represented in \Fref{fig:figSIV}. }
\label{fig:fighop} 
\end{figure}

A simple extrapolation of this picture would forecast a progressively more substantial degradation of the conductance with further stretching. However, inspection of \Fref{fig:figTE}a reveals otherwise: resonant transmission is revived beyond a geometry-controllable threshold when the system enters stage 2 of deformation. At 25.1\,\% strain, resonant transmission is recovered at particular energies within the lowest miniband, and further stretching to 34.7\,\% strain, rather than degrading, further improves this situation with resonant peaks re-appearing again throughout the entire lowest miniband as in the undeformed device. As expected, the width of each miniband is now smaller than in the undeformed state, a combined effect of the increase in inter-dot distance, and the decrease of electronic hoppings in the carbon lattice. \Fref{fig:figLDOS}c exhibits the LDOS for one of the revived conductance peaks of the same system that shows a strong similarity with the undeformed counterpart in \Fref{fig:figLDOS}a.

It is clear from both the conductance and the LDOS that deforming the system well inside stage 2 restores the coupling between the localized states trapped at the internal mini-edges, that is strongly affected during stage 1. As pointed out earlier, the major difference between stages 1 and 2 is that, in the latter, further elongation proceeds through deformation of the C-C bonds, and therein lies the mechanism that promotes the enhanced coupling and the revival of resonant transmission.
As one intuitively anticipates, and can be explicitly seen in \Fref{fig:figSIV}b, the deformations in question are borne almost entirely by the bonds along the segments that link neighboring units of the kirigami.   A comparison of the bond length distribution in stage 1 (\Fref{fig:figSIV}a) versus stage 2 (\Fref{fig:figSIV}b) indeed shows that the first regime is characterized by small and seemingly random variations, whereas the latter features a clear and periodic pattern of deformation hot spots where the C-C bond is stretched in excess of 20\,\%. 

Variations in bond-length imply perturbations to the electronic hopping [$t_{ij}$, \eqref{eq:tij}], which is the quantity that directly affects the transport. However, it is important to mention that the C-C bond map excludes bending contributions that should dominate the hopping modifications in stage 1. To clarify this point, we plot in \Fref{fig:fighop} the corresponding hopping distribution maps. The variations are of the order of $\sim\pm 10$\,\% and $\sim\pm 20$\,\% at 15.5\,\% and 34.7\,\% elongation, respectively. More than the magnitude, the key difference lies in the sign and spatial pattern of these hopping variations: in the first case, they appear randomly distributed and one sees places of both enhanced and reduced hopping compared to the undeformed structure; in the second, the pattern is clearly periodic and dominated by reduced hopping ($\sim 20$\,\%) spatially correlated with the strain hot spots of \Fref{fig:figSIV}(b).
These electronic ``weak links'' reinforce the super-periodicity of the structure, while simultaneously contributing to localize further the electronic states within each segment, thereby promoting the revival of the miniband and stop-gap structure in the conductance.

In summary, the profile of the conductance at low energies in these graphene structures is typical of a system of weakly coupled quantum dots: minibands of resonant transmission alternating with stop-gaps due to the superlattice periodicity along the longitudinal direction. Direct inspection of the LDOS at any of the resonant energies provides a direct confirmation of this in real-space. Analysis of the evolution of the conductance and the LDOS with deformation shows that the resonant tunneling assisted by localized states is strongly suppressed during stage 1 of deformation. The local perturbations to the electronic hopping in this regime are sufficient to disturb the overlap between the localized states. This effect is reversed in stage 2 by the appearance of strong strain barriers spanning the length of the system along the links between individual elements of the periodic structure. These have a confining nature from the electronic point of view, which consistently explains the reinforcement of overlap between localized states, and the revival of resonant transmission, at the highest deformations.

\section{I-V characteristics and negative differential resistance}

\begin{figure}[tb]
\vspace{-0.2cm}
\centerline{\includegraphics[width=0.3\textwidth]{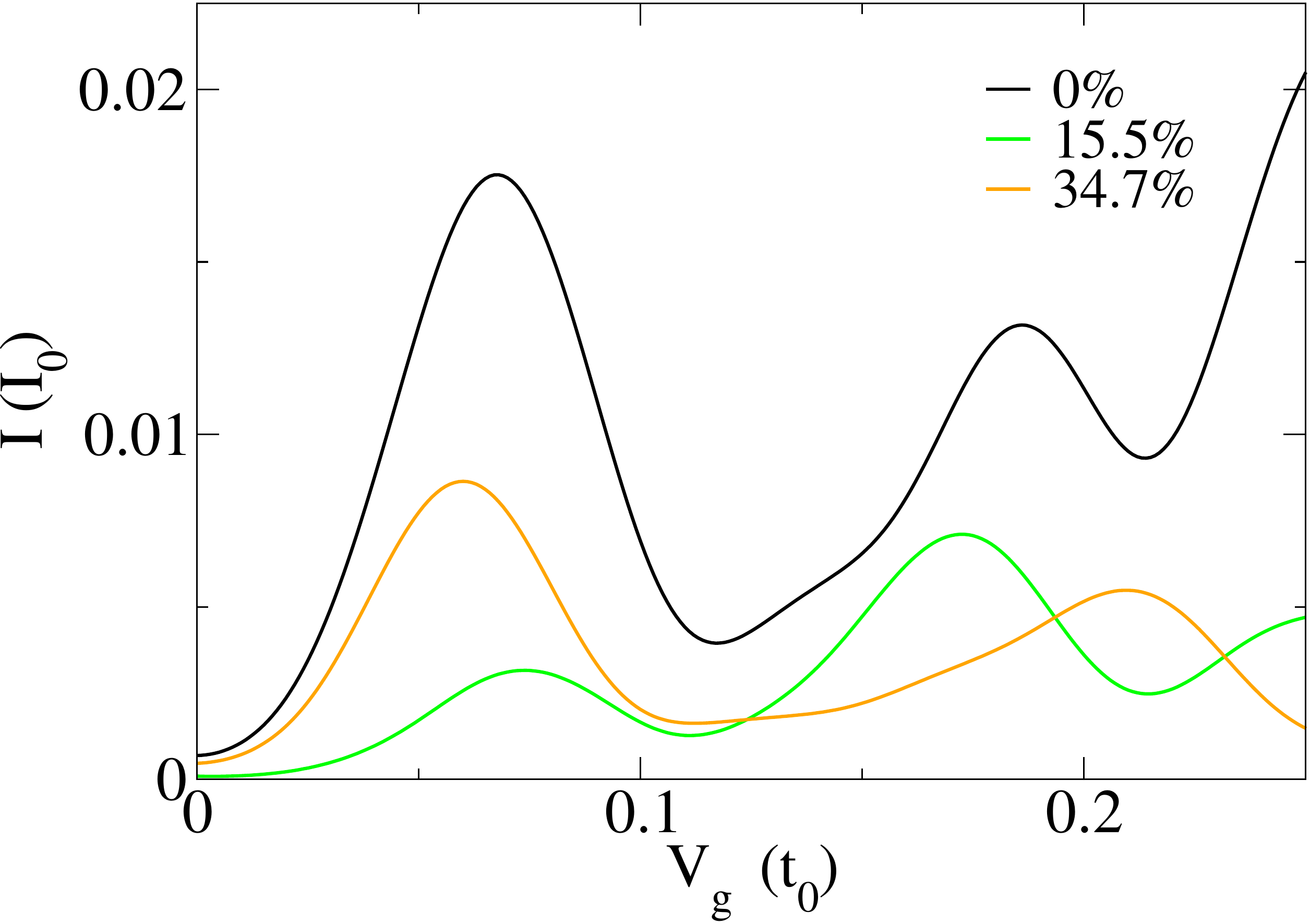}}
\vspace{-0.2cm}
\caption{Current \textit{vs} gate voltage for the kirigami deformed by 0\,\%, 15.5\,\%, and 34.7\,\%. We used a bias voltage of $V_{SD} = 100$\,meV, the current is expressed in units of $I_0 = 2e/h$, and the gate voltage in units of $t_0$. 
}
\label{fig:fig_iv}
\end{figure}

In principle, to fully  examine the current-voltage characteristics of these kirigami we should calculate the transmission function for different bias and gate voltages, according to \Eqref{eq:Itot}. However, similar to the experimental setup used in reference \onlinecite{Blees:2015fk} where the kirigami was immersed in an electrochemically-controlled liquid gate with a fixed bias voltage $V_{SD} = 100$ meV, we restrict our analysis to the electric current as a function of gate voltage for a fixed bias. \Fref{fig:fig_iv} displays current for kirigami deformed by 0, 15.5, and 34.7\,\% under a fixed $V_{SD} = 100$\,meV, and at $T=300$\,K. 

One can see a clear oscillation of the current with increasing $V_g$, a behavior 
that traces back to the underlying mini-band structure of the conductance 
discussed in the context of \Fref{fig:figTE}. The fact that below 
$V_g=0.1\,t_0$ the current magnitude first decreases with deformation, but then 
increases again in stage 2, is a natural consequence of the revival of resonant 
transmission at low energies promoted by the localized strain that emerges in stage 2. The most interesting aspect of the I-V characteristics shown 
in \Fref{fig:fig_iv} is that the current oscillation immediately implies that 
these kirigami have negative differential resistance over reasonably large 
intervals of $V_g$, which is a much sought-after property for new classes of 
nonlinear electronic devices \cite{5391729,1073171}.

\section{Introducing dephasing}

\begin{figure}[t]
\vspace{-0.2cm}
\centerline{\includegraphics[width=0.45\textwidth]{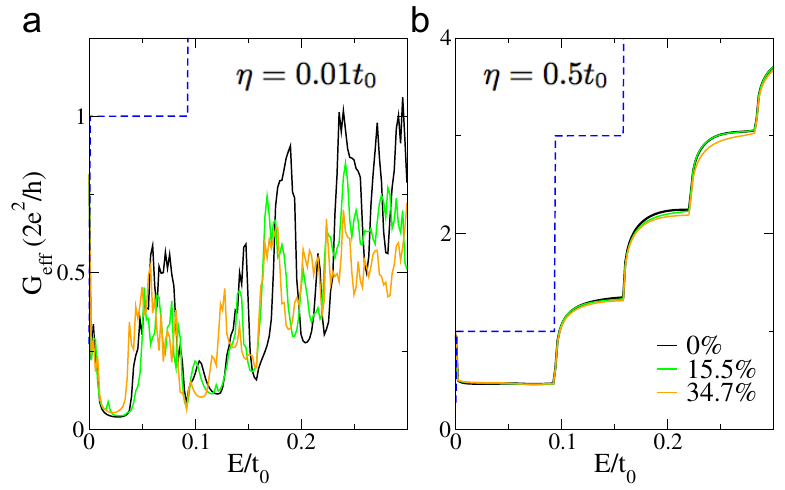}}
\vspace{-0.2cm}
\caption{ Effective conductance in the presence of (a) weak ($\eta = 0.01t_0$, $\tau_{\phi} \approx 6.1\times 10^{-15}$s) and (b) strong ($\eta = 0.5t_0$, $\tau_{\phi} \approx 1.2\times 10^{-16}$s) phase-breaking 
scatterers for different deformations. The dashed blue line corresponds to the conductance of a pristine zigzag nanoribbon with the same width of our contacts. Be aware of the different scales in the vertical axes. }
 \label{fig:figdph}
\end{figure}

So far, our electronic transport model considers coherent transport, including only edges and mechanical deformations as source of scattering. To address the robustness of the strain modulation of the conductance in a more realistic scenario, we now turn to effects of random scatterers modeled within the dephasing B\"uttiker-probe model \cite{Buttiker4terminals}. To achieve that, we distribute voltage probes as phase-breaking scatterers all over the kirigami (except the left and right edges), and set the current at each probe to zero. Given that we are not interested in the particular physical mechanism (impurities, phonons or  electrons) behind the destruction of quantum coherence, we fix the self-energy of each  B\"uttiker-probe to $\Sigma_{\phi} = -i\eta$ \cite{PhysRevB.75.081301}, which is related to the phase relaxation time by $\tau_{\phi} = \hbar/2\Gamma_{\phi} = \hbar/4\eta$ \cite{Datta,Pastawski}. In \Fref{fig:figdph}, we plot the effective conductance that obtains for weak ($\eta = 0.01t_0$, $\tau_{\phi} \approx 6.1\times 10^{-15}$s) and strong ($\eta = 0.5t_0$, $\tau_{\phi} \approx 1.2\times 10^{-16}$s) phase-breaking under various deformations. Inspecting \Fref{fig:figdph}a, we can see that weak dephasing is sufficient to destroy the resonant transmission and mini-band structure in the transmission, irrespective of the amount (or absence) of deformation. Strong phase-breaking processes (\Fref{fig:figdph}b) completely destroy any fine structure in the conductance associated with the particular geometry of the device that are still visible in \Fref{fig:figdph}a. In this case, there are no differences among the effective conductances of unstrained and deformed kirigami, all of which display the step-wise behavior characteristic of a pristine graphene nanoribbon, with the steps appearing at energies that are determined by the overall width of the device (parameter $b$ in \Fref{fig:structure}a).

To illustrate the distinction between the weak and strong dephasing regimes used in \Fref{fig:figdph}, it is instructive to compute the associated phase-relaxation lengths, $l_{\phi} = \sqrt{D\tau_{\phi}}$ \cite{Datta} using the representative carrier diffusion coefficient $D=0.3~\text{m}^2/\text{s}$ \cite{Berger:2006cr}. 
In the case of weak dephasing discussed above the phase-relaxation length is larger than the device size, $l_{\phi} \approx 42\,\text{nm} > L$, whereas that value for strong dephasing is similar to the periodicity of the kirigami, 
$l_{\phi} \approx 6\,\text{nm} \approx w$. Coherent effects in the conductance are observed when the typical device dimensions are smaller than the phase relaxation length \cite{Datta,PhysRevB.75.081301,Berger:2006cr}. In our case, the formation of gaps and minibands is a coherent effect created by destructive and constructive interference of electron waves at the periodically repeating segments of the kirigami (the localized states arise as the electron bounces back and forth without losing its phase within different sections of the kirigami). \Fref{fig:figdph}b shows a conductance that has become insensitive to any effect associated with strain or even the geometry. This confirms that, in order to observe efficient mechanical control over the conductance, the phase-relaxation length must be larger than the segments of the kirigami that harbor the localized states that support the resonant tunneling at low energy. 
Measurements of the phase-relaxation length at low temperature in graphene have found $l_{\phi} \approx 1\,\mu\text{m}$ \cite{Berger:2006cr}, which is much smaller than the size of the typical internal lengths used in the kirigami experimentally probed by Blees and collaborators \cite{Blees:2015fk}. This might partially explain why deformed kirigami used in that experiment did not show noticeable sensitivity of the conductance to the elongation.

\section{Final remarks}

Our study shows that, when the typical feature sizes of a graphene-based kirigami are in the nanoscale, the electric conductance at low energies might be governed by resonant tunneling through states that are localized by the specific local geometry of each repeating element of the kirigami. 
The longitudinal periodicity of the structure results in an efficient coupling (strong overlap) between these states in the undeformed configuration. This state of affairs is, however, strongly sensitive to stage 1 deformations, and the conductance easily degrades in that regime where the overall elongation is a result of the structural twisting and bending, rather than extensive stretching of the C-C bonds. The electronic overlap is reinforced for stage 2 deformations as a result of the confining nature of the localized strain barriers that set in during this stage, and resonant transmission is hence revived at high overall elongations.
The regime of resonant tunneling exists within well defined energy minibands isolated from each other by sizable transmission stop-gaps. Their existence results in a strong oscillation of the I-V characteristic as a function of gate voltage, and opens the possibility of driving the system from a conventional resistive regime to one of negative differential resistance, by simple electrostatic gating.

To explain the evolution of the conductance profile with elongation we analyzed directly the bond stretching and the perturbations to the nearest-neighbor hopping. It is customary in discussions of strained graphene to introduce the concept of pseudomagnetic field (PMF) \cite{Kane:1997,Suzuura:2002,Vozmediano2010Gauge} and map strain fields to PMFs in order to, for example, obtain a semi-classical intuition about how certain strain patterns disturb the motion of electrons \cite{PhysRevB.84.081401}. In the present case, however, the characteristic dimensions of our device are small and, in addition, the strain distribution displays sharp variations within these small scales. This restricts the usefulness of the PMF concept and any semi-classical picture to interpret the effects of strain. We therefore focused directly in the changes affecting the nearest-neighbor hopping amplitudes since these are the quantities that more direct and fundamentally determine the conductance.

\begin{figure}[t]
  \centerline{\includegraphics[width=0.45\textwidth]{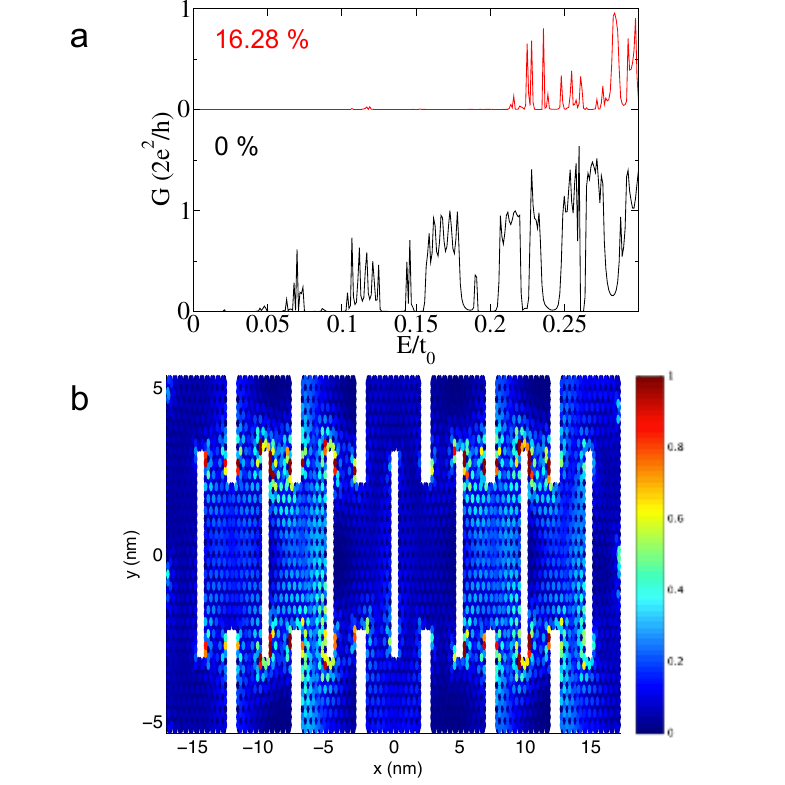}}
  \caption{ (a) Conductance of the armchair kirigami deformed by 0\,\% and 16.28\,\%. (b) LDOS for the undeformed armchair kirigami at $E=0.121t_0$ }
  \label{fig:armchair}
\end{figure}

Finally, in our main discussion we focused on a kirigami whose horizontal segments consisted of zig-zag strips of graphene. Not surprisingly, the mini zig-zag edges parallel to the transverse direction play an important role in stabilizing the localized states in the undeformed structure that are key for the conductance profile at low energy. It is important to reiterate that this is not a limitation, for \emph{the key physical ingredient is the existence of localized states defining local quantum dots}, and the interplay between the strain-induced changes of the C-C hopping and effective inter-dot coupling upon mechanical stretching. To be specific, in \Fref{fig:armchair}a we show that rotating the underlying lattice by 90 degrees in the same kirigami leads to a similar profile of mini-band and stop-gap low-energy conductance, and that it is equally sensitive to deformation: for example, at about 16\,\% elongation the miniband structure is entirely suppressed. In a future study we shall undertake the impact of edge roughness. We can, however, anticipate that edge roughness is expected to \emph{improve} the scenario we describe here because this type of disorder promotes further localization of low-energy states in narrow graphene structures \cite{Todd2008Quantum,PhysRevLett.104.056801}, and that explains the experimentally observed gaps and Coulomb blockade in rough (lithographically patterned) graphene nanoribbons  \cite{Sols2007Coulomb,Todd2008Quantum,Stampfer2009Energy,Gallagher2010Disorderinduced}. We therefore predict that, in a realistic scenario, individual segments of a kirigami can behave as true quantum dots with a coupling (and, consequently, an overall transmission) amenable to modulation through the same type of deformation discussed here. The conditions for such behavior are expected to be very encompassing, depending only on choosing appropriate combinations of scales and geometries capable of hosting localized quantum-dot states, supported either by the geometry, disorder, or interactions. 

\acknowledgments

DAB acknowledges support from FAPESP grant 2012/50259-8. ZQ acknowledges the support of the Mechanical Engineering and Physics Departments at Boston University. VMP was partly supported by the National Research Foundation (Singapore) under its Medium-Sized Centre programme.

\bibliographystyle{apsrev}
\bibliography{bibkirigami}

\end{document}